\documentclass[twocolumn]{aastex6}

\shorttitle{Chandra HETG spectrum of 3C~390.3}
\shortauthors{Tombesi et al.}

\begin{document}


\title{The complex circumnuclear environment of the broad line radio galaxy 3C~390.3 revealed by Chandra HETG}


\author{F. Tombesi$^{1,2}$, J.~N. Reeves$^{3,4}$, T. Kallman$^{1}$, C.~S. Reynolds$^2$, R.~F. Mushotzky$^2$, V. Braito$^5$, E. Behar$^{2,6}$, M.~A. Leutenegger$^1$ and M. Cappi$^7$}
\affil{$^{1}$X-ray Astrophysics Laboratory, NASA/Goddard Space Flight Center, Greenbelt, MD 20771, USA; francesco.tombesi@nasa.gov}
\affil{$^{2}$Department of Astronomy, University of Maryland, College Park, MD 20742, USA; ftombesi@astro.umd.edu}
\affil{$^{3}$Center for Space Science and Technology, University of Maryland Baltimore County, 1000 Hilltop Circle, Baltimore, MD 21250, USA}
\affil{$^4$Astrophysics Group, School of Physical and Geographical Sciences, Keele University, Keele, Staffordshire, ST5 5BG, UK}
\affil{$^5$INAF - Osservatorio Astronomico di Brera, Via Bianchi 46 I-23807 Merate (LC), Italy}
\affil{$^6$Department of Physics, Technion 32000, Haifa 32000, Israel}
\affil{$^7$INAF-IASF Bologna, Via Gobetti 101, I-40129 Bologna, Italy}



\begin{abstract}


We present the first high spectral resolution X-ray observation of the broad-line radio galaxy 3C~390.3 obtained with the high energy transmission grating (HETG) spectrometer on board the \emph{Chandra} X-ray Observatory. The spectrum shows complex emission and absorption features in both the soft X-rays and Fe K band. We detect emission and absorption lines in the energy range between E$=$700--1000~eV associated with ionized Fe L transitions (Fe~XVII--XX). An emission line at the energy of E$\simeq$6.4~keV consistent with the Fe K$\alpha$ is also observed. Our best-fit model requires at least three different components: (i) a hot emission component likely associated with the hot interstellar medium in this elliptical galaxy with temperature $kT$$=$$0.5\pm0.1$~keV; (ii) a warm absorber with ionization parameter log$\xi$$=$$2.3\pm0.5$~erg~s$^{-1}$~cm, column density log$N_H$$=$$20.7\pm0.1$~cm$^{-2}$, and outflow velocity of $v_{out}$$<$150~km~s$^{-1}$; (iii) a lowly ionized reflection component in the Fe K band likely associated with the optical broad line region or the outer accretion disk. These evidences suggest the possibility that we are looking directly down the ionization cone of this active galaxy and that the central X-ray source only photoionizes along the unobscured cone. This is overall consistent with the angle-dependent unified picture of active galactic nuclei.

\end{abstract}




\section{Introduction} \label{sec:intro}

It is becoming clear that supermassive black holes (SMBHs) at the center of active galactic nuclei (AGN) may play an important role in galaxy evolution through a process known as ``feedback'' (e.g., Silk \& Rees 1998; Fabian 1999; Veilleux et al.~2013; Tombesi et al.~2015; Nardini et al.~2015). Depending whether it is mediated by winds or jets it is commonly referred to as ``quasar'' or ``radio'' mode feedback (e.g., Fabian 2012; King \& Pounds 2015). However, several fundamental questions are still open: what is the origin of the difference between feedback by radio-quiet and radio-loud AGN? Why and how is the radio-quiet and radio-loud phase linked to the galaxy type and large-scale environment? Is feedback different for isolated, group or cluster galaxies? The investigation of the complex environment surrounding AGN may shed light on some of these fundamental issues (e.g., Ineson et al.~2015). 

Absorption from layers of photoionized gas in the circumnuclear regions of AGN is commonly observed in Seyfert galaxies and quasars. This material can be observed in the X-ray spectra through blue-shifted absorption lines from various elements over a wide range of ionizations, column densities, and velocities. The parameters of these winds range from log$\xi$$\simeq$0--4~erg~s$^{-1}$~cm, $N_\mathrm{H}\simeq 10^{20}$--$10^{22}$~cm$^{-2}$ and $v_{out}$$\sim$100--1000~km~s$^{-1}$ for the so-called ``warm absorbers'' (WAs) (e.g., Halpern 1984; Nandra \& Pounds 1992; Reynolds 1997; Crenshaw \& Kraemer 2012; Tombesi et al. 2013a; Kaastra et al.~2014) to more extreme values of log$\xi$$\simeq$4--6~erg~s$^{-1}$~cm, $N_\mathrm{H}\simeq 10^{22}$--$10^{24}$~cm$^{-2}$ and $v_{out}$$\sim$10,000--100,000~km~s$^{-1}$ for the so-called ``ultrafast outflows'' (UFOs) (e.g., Tombesi et al.~2010a; Gofford et al. 2013). The ionization parameter is defined as $\xi$$=$$L_\mathrm{ion}/n r^2$ erg~s$^{-1}$~cm (Tarter, Tucker \& Salpeter 1969) where $L_\mathrm{ion}$ is the ionizing luminosity between 1 Ryd and 1000 Ryd (1 Ryd $=$ 13.6 eV), $n$ is the number density of the material, and $r$ is the distance of the gas from the central source. Their origin and acceleration mechanisms of these winds are still debated (Fukumura et al. 2010, 2015; King \& Pounds 2014), but they seem to be part of a common large-scale outflow originating from the accretion disk and/or torus (e.g., Tombesi et al.~2013a). 


\begin{deluxetable}{lcccccccc}
\tablecaption{Chandra HETG observations log. \label{tab:table}}
\tablehead{
\colhead{Obs} & \colhead{ID} & \colhead{Date} & \colhead{Exp} & \colhead{Rate}
}
\startdata
1 & 16531 & 2014/6/15 & 50 & 0.79/0.38\\
2 & 16220 & 2014/6/18 & 50 & 0.70/0.34\\
3 & 16530 & 2014/7/12 & 50 & 0.85/0.40\\
\enddata
\tablecomments{Columns: observation number; observation ID; observation date; exposure in ks; MEG/HEG count rates.}
\end{deluxetable}

In stark contrast, the X-ray evidence for winds was scarce in radio-loud galaxies with powerful relativistic jets. Recent sensitive observations of broad-line radio galaxies (BLRGs) have started to subvert this view. The detection of WAs was reported in 3C~382, 3C~445, 3C~390.3 and 4C$+$74.26 (Ballantyne 2005; Reeves et al.~2009, 2010; Torresi et al. 2010, 2012). Moreover, UFOs have been reported in 3C~111, 3C~120, 3C~390.3, 3C~445, 4C$+$74.26 and Cygnus A (Tombesi et al.~2010b, 2011; Ballo et al.~2011; Braito et al.~2011; Gofford et al.~2013, 2015; Reynolds et al.~2015). A recent X-ray study of a sample of 26 radio galaxies reported that the frequency of UFOs is likely in the interval $f$$\simeq (50\pm20)$\% (Tombesi et al.~2014). Thus, contrary to the jet dichotomy, it seems that winds may be present in both luminous radio-quiet and radio-loud AGN.  

Indeed, from a theoretical point of view, winds are expected in radio-loud AGN as ingredients for jet formation (Blandford \& Payne 1982) and the relativistically moving jet plasma should be enveloped in a sub-relativistic wind, possibly helping the initial collimation of the jet (McKinney 2006; Fukumura et al. 2014). The presence of both the accretion disk, winds and jets in radio galaxies makes them the ideal objects to study the interplay among these components (e.g., Marscher et al. 2002; Chatterjee et al. 2009, 2011; Tombesi et al.~2011, 2012, 2013b; Lohfink et al.~2013). 

Moreover, spectral features indicating reflection from the inner accretion disk and the parsec scale torus or broad line region have been identified in some high signal-to-noise spectra of radio galaxies. They appear to be similar but weaker than in Seyfert galaxies, possibly indicating a different disk ionization state, lower column densities, lower covering fraction or a different illumination from the base of the jet (e.g., Sambruna et al.~2009; Grandi \& Palumbo 2007; Kataoka et al. 2007; Tombesi et al.~2013b; Tazaki et al.~2013; Bostrom et al. 2014). 

The dichotomy between radio-quiet and radio-loud AGN is still not fully understood, but it seems that the latter host more massive black holes, and are more preferentially found in elliptical galaxies and mergers (e.g., Chiaberge et al.~2011, 2015). One related possibility is also that the AGN is connected through feeding and feedback with the large-scale environment and, therefore, both the black hole and the host galaxy are linked to each other. For instance, elliptical galaxies are known to possess a hot interstellar medium (ISM) with temperature of the order of $kT$$\sim$0.1--1~keV (Werner et al.~2009; Kim \& Pellegrini 2012). Dozens or more soft X-ray emission lines, particularly a forest of Fe L lines, is expected to provide the main cooling mechanism (e.g., Xu et al.~2002; Werner et al.~2009). The physical properties of such hot ISM may be directly related to the formation and evolution of these systems, via star formation episodes, the passive evolution of their aging stellar population, environmental effects such as stripping, infall, and mergers, and the growth of supermassive black holes (see Kim \& Pellegrini 2012 for a review).

For the first time, we show the analysis 150~ks \emph{Chandra} HETG observation of the BLRG 3C~390.3 ($z = 0.0561$). The unique capabilities of the \emph{Chandra} HETG of combining high energy resolution and high sensitivity in the wide E$=$0.5--7~keV band are crucial for the detection of emission and absoprtion lines from a wide range of ionization species. 

\section{Data reduction and analysis} 

The \emph{Chandra} HETG observation of 3C~390.3 is composed of three exposures performed within one month in 2014 for a total of 150~ks, see Table~1 for details. The spectra were extracted using the \emph{CIAO} package v4.7 and the associated CALDB (updated to March 2015). Only the first order dispersed spectra are considered for both the MEG (Medium Energy Grating) and HEG (High Energy Grating) and the $\pm$1 orders for each grating were subsequently combined for each sequence. No significant spectral variability is observed between the three exposures and the spectra are consistent with only minor $\simeq$4\% variations in source count rate. Therefore, the spectra were combined from all three sequences to yield a single 1st order spectrum for each of the MEG and HEG. The background count rate is negligible. The resultant spectra were subsequently binned to $\Delta\lambda = 0.023$ \AA\, and $\Delta\lambda = 0.012$ \AA\, bins for MEG and HEG respectively, which match their full width half maximum (FWHM) spectral resolution. The MEG and HEG spectra were analyzed in the observer frame energy intervals E$=$0.5--7~keV and E$=$1--7.5~keV, respectively. The analysis of the background subtracted source spectra was performed using \emph{XSPEC} v.12.8.2 and the C-statistic was applied. We perform a simultaneous fit of the MEG and HEG spectra including a free cross-normalization constant, which resulted being within 5\%. All parameters are given in the rest frame of the source at $z=0.0561$ and the errors are at the 1$\sigma$ level if not otherwise stated. In all the fits, a Galactic absorption of $N_H = 4\times 10^{20}$~cm$^{-2}$ was adopted (Kalberla et al.~2005).

  \begin{figure}
  \centering
   \includegraphics[width=8cm,height=6cm,angle=0]{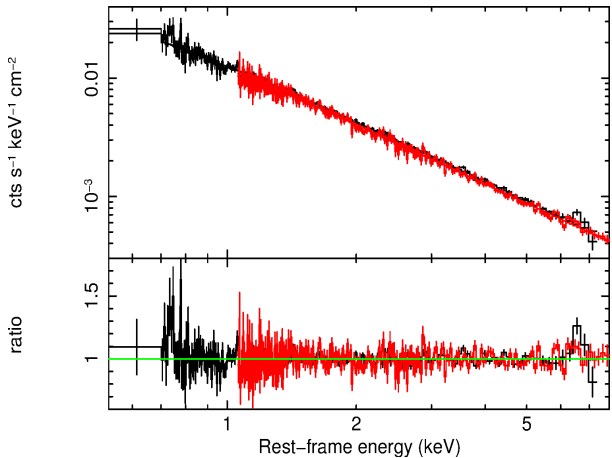}
   \caption{Combined \emph{Chandra} MEG (black) and HEG (red) spectra of 3C~390.3 compared to a Galactic absorbed power-law continuum model. \emph{Upper panel:} spectra and the continuum model. \emph{Lower panel:} data to model ratios with respect to the continuum model. The data are binned to 4$\times$ the FWHM resolution and to a minimum signal-to-noise of 5 for clarity.}
    \end{figure}

\subsection{Phenomenological spectral analysis}

We started the spectral analysis including a continuum power-law with $\Gamma$$\simeq$1.7 ($C/\nu = 9166/8898$). We checked that a neutral absorption component intrinsic to the source is not required. A possible soft excess component is also not required by the data. An inspection of the data in Fig.~1 shows the possible presence of emission and absorption residuals in both the soft X-ray and Fe K bands

In the upper panel of Fig.~2 we observe three emission features in the energy range between E$=$0.7--0.85~keV. 
The energy and width of the first emission line at E$\simeq$734~eV suggest an interpretation\footnote{The line identifications are derived from the National Institute of Standards and Technology (NIST) database or Verner et al.~(1996).} as the doublet Fe~XVII 2p$\rightarrow$3s at E$=$727~eV and E$=$739~eV, respectively. We note that the energy of this feature is also consistent with the O~VII RRC at E$=$739~eV (e.g., Liedahl \& Paerels 1996). However, this latter interpretation is less likely because there are no other lines detected in the present spectrum that can be associated with O~VII. An [O~VII] forbidden line at 560~eV was reported in the \emph{XMM-Newton} Reflection Grating Spectrometer (RGS) and \emph{Suzaku} spectra (Sambruna et al.~2009; Torresi et al.~2012). The very limited signal-to-noise of the MEG at that energy does not allow us to detect such a feature if present. 

We also note another emission line at the energy of E$\simeq$779~eV, which can be associated with the Fe~XVIII 2p$\rightarrow$3s doublet at E$=$777~eV and E$=$779~eV, respectively. Although, a possible partial contamination from O~VIII Ly$\beta$ at E$=$774~eV can not be fully excluded (e.g., Kinkhabwala et al.~2002; Sako et al.~2002). However, we note that the apparent absence of a related strong O~VIII Ly$\alpha$ line seem to rule out a significant contribution from O~VIII Ly$\beta$. Finally, there is indication of a fainter emission feature at E$\simeq$808~eV which can be associated with the Fe~XVII 2p$\rightarrow$3d doublet at E$=$802~eV and E$=$812~eV, respectively. Therefore, the soft X-ray emission lines are dominated by Fe L-shell transitions. The best-fit parameters of these lines are reported in Table~2.

  \begin{figure}
  \centering
   \includegraphics[width=7.5cm,height=5.7cm,angle=0]{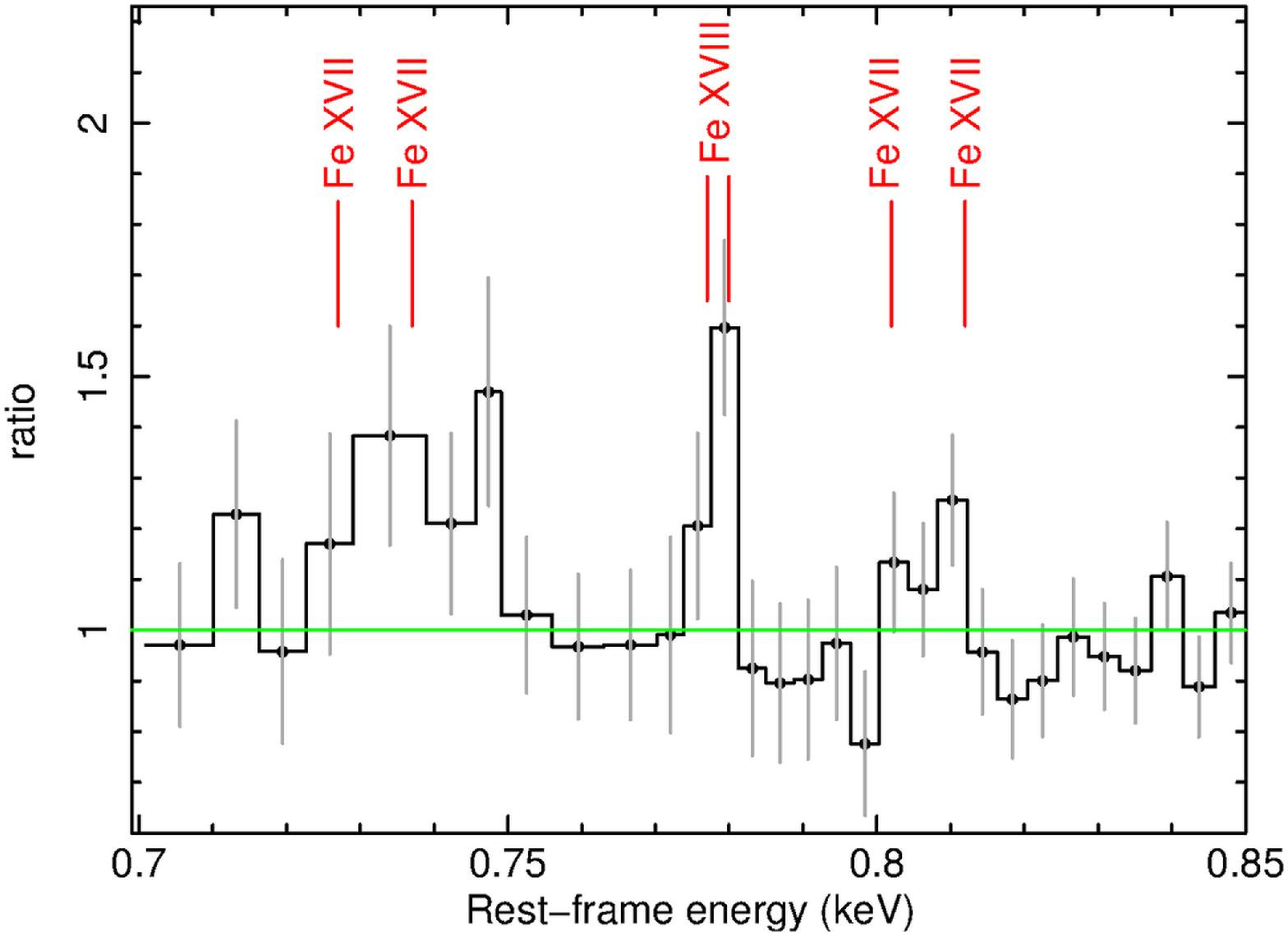}
   \includegraphics[width=7.5cm,height=5.7cm,angle=0]{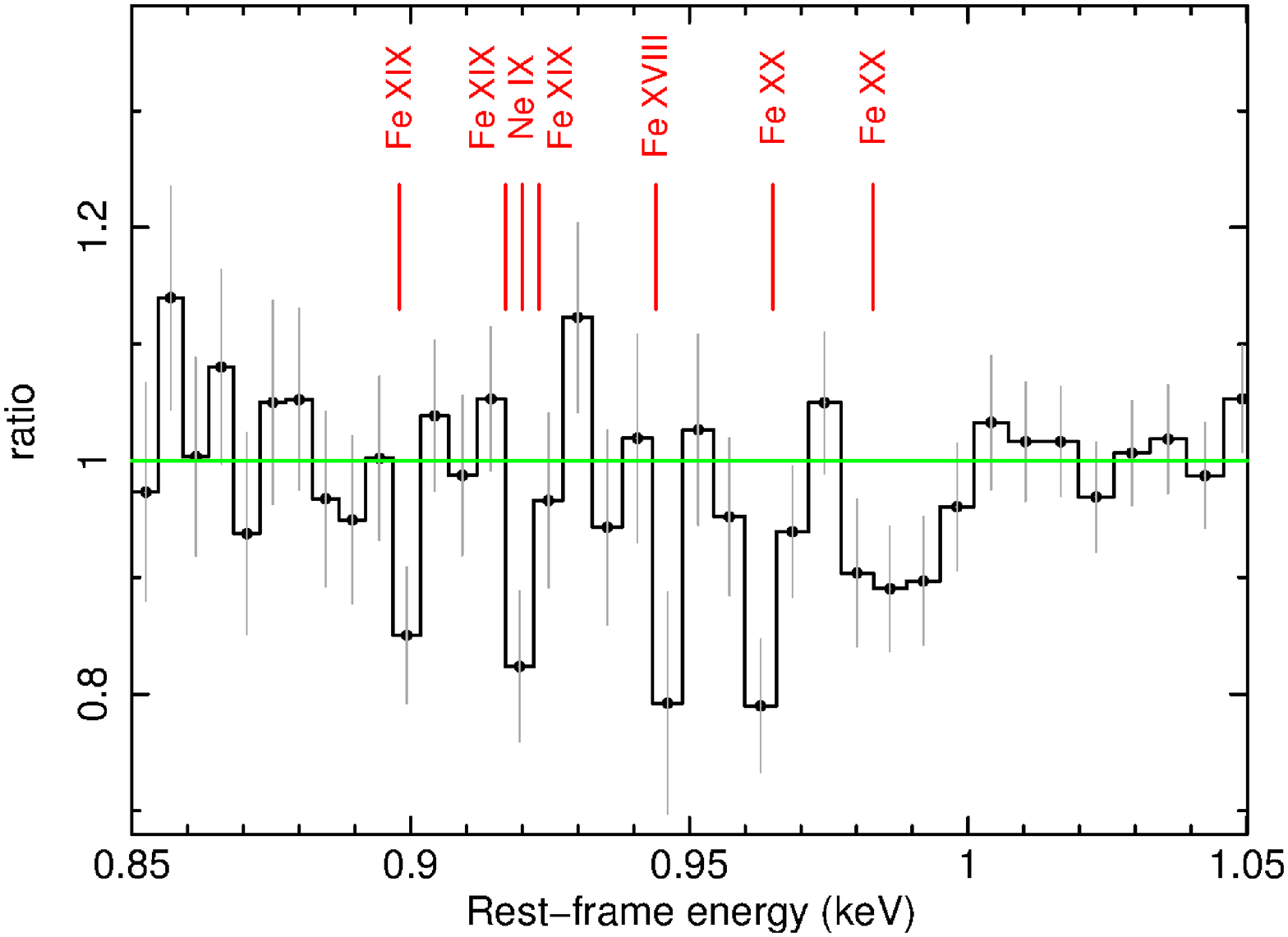}
   \includegraphics[width=7.5cm,height=5.7cm,angle=0]{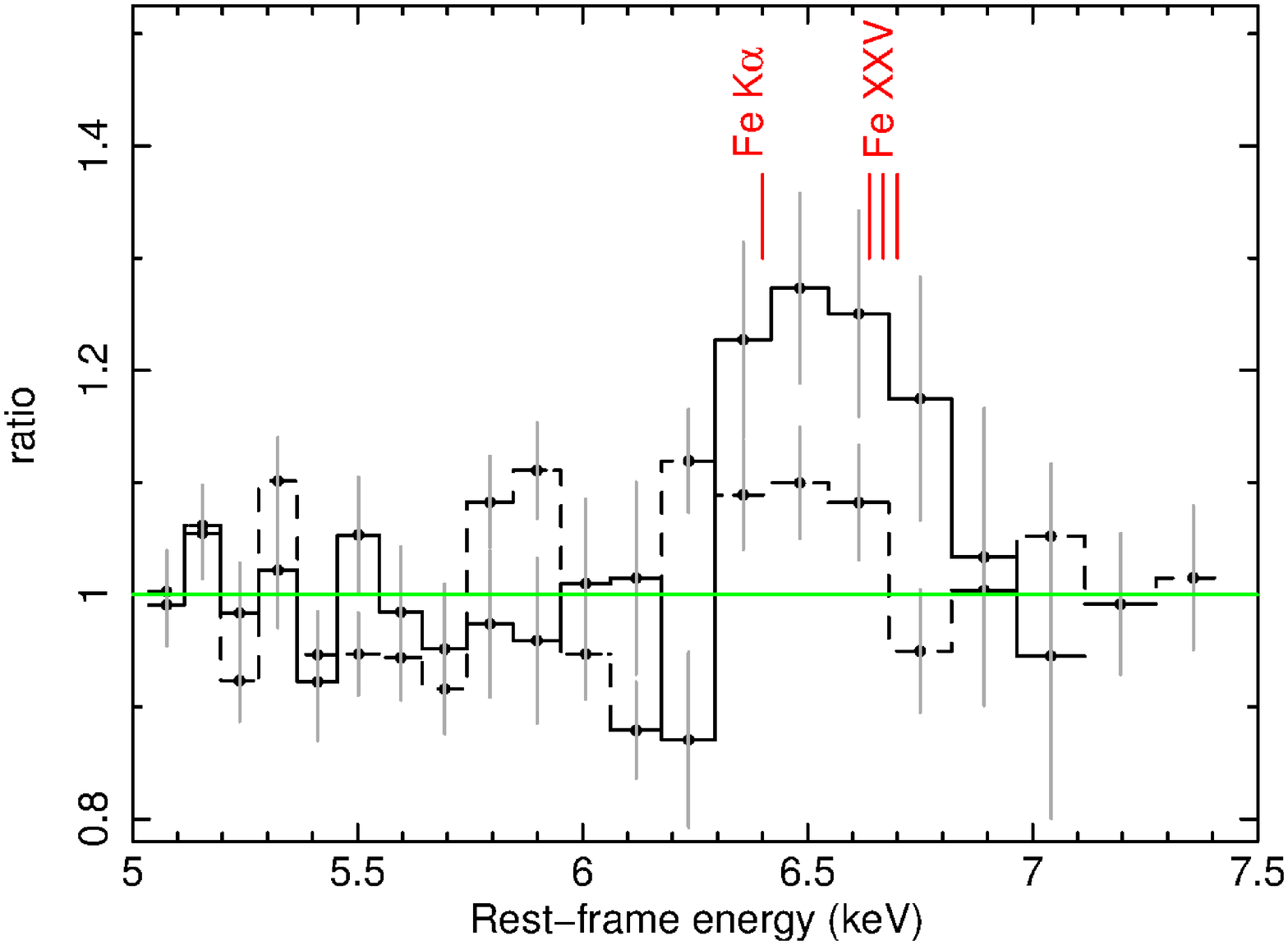}
   \caption{Data to model ratios of the \emph{Chandra} MEG (solid line) and HEG (dashed line) spectra of 3C~390.3 with respect to a Galactic absorbed power-law continuum model. \emph{Upper panel:} soft X-ray emission lines in the E$=$0.70--0.85~keV energy band. \emph{Middle panel:} soft X-ray absorption lines in the E$=$0.85--1.05~keV energy band. \emph{Lower panel:} Fe K emission lines in the E$=$5.0--7.5~keV energy band. The data are binned to 4$\times$ the FWHM resolution and a signal-to-noise of 5 for better clarity.}
    \end{figure}

Five absorption features are observable in the energy range between E$=$0.85--1.05~keV in the middle panel of Fig.~2. The lines are likely associated with these Fe L-shell transitions: Fe~XIX 2p$\rightarrow$3d at E$=$899~eV, E$=$917~eV, and E=$=$923~eV; Fe~XVIII 2s$\rightarrow$3p at E$=$942~eV; Fe~XX 2p$\rightarrow$3d at E$=$967~eV; and Fe~XX 2p$\rightarrow$3d at E$=$981~eV. We note that the absorption at E$=$922~eV may also be consistent with the position of the resonance absorption of Ne~IX at E$=$920~eV. This possibility is discussed in section 2.1.1. The best-fit parameters of the lines are reported in Table~2. We show only the lines with a fit improvement of $\Delta C \ge 4$ for two additional degrees of freedom, corresponding to a significance of higher than 90\%. We do not list the line at E$\simeq$900~eV in Table~2 because it provides an improvement of $\Delta C = 3$. Four of the absorption lines are marginally resolved, with velocity widths in the range between $\sigma_v$$\sim$500--1500 km~s$^{-1}$. The presence of Fe~XX absorption was reported also in a short \emph{XMM-Newton} RGS spectrum obtained in 2004 (Torresi et al.~2012).

Two emission features are observable in the Fe K band in the lower panel of Fig.~2. The most intense line at E$\simeq$6400~eV is likely associated with the neutral or lowly ionized Fe K$\alpha$ fluorescent line. The line is resolved, with a width of $\sigma$$=$$90\pm33$~eV. An additional fainter feature is observed at E$\simeq$6640~eV, which is likely associated with the Fe~XXV 2s$\rightarrow$2p forbidden line at E$=$6637~eV. The best-fit parameters are reported in Table~2.

A broad ($\sigma$$\simeq$500~eV) and intense (EW$\simeq$70~eV) emission line at E$\simeq$6.6~keV, possibly associated with blurred disk reflection from the accretion disk, was reported in the previous \emph{XMM-Newton} and \emph{Suzaku} spectra (Sambruna et al.~2009). The limited signal-to-noise of the HETG data in the Fe K band allows to place only an upper limit of 250~eV on the line broadening. The limited sensitivity of the HETG at E$\ge$7~keV also does not allow to observe absorption lines from an ultrafast outflow if present. An ultrafast outflow was detected in the \emph{Suzaku} spectrum at E$=$8.1~keV, which is outside the available energy range of the HETG (Tombesi et al.~2010b; Gofford et al.~2013).

\floattable
\begin{deluxetable*}{cccccc}[t!]
\tablecaption{Best-fit parameters of the emission and absorption lines.}
\tablehead{
\colhead{E} & \colhead{$\sigma$} & \colhead{$I$} & \colhead{EW} &
\colhead{ID} & \colhead{$\Delta$C}
}
\startdata
\hline\hline
\multicolumn{6}{c}{Soft X-ray emission lines}\\
\hline
$734\pm4$ & $14\pm5$ & $30^{+6}_{-9}$ & $11.0\pm3.0$ & Fe~XVII 2p$\rightarrow$3s & 19\\[+4pt]
$779\pm1$ & $2.0\pm0.6$ & $6\pm3$ & $2.5\pm0.8$ & Fe~XVIII 2p$\rightarrow$3s& 13\\[+4pt]
$808\pm1$ & $<4^a$ & $3\pm1$ & $1.4\pm0.6$ & Fe~XVII 2p$\rightarrow$3d & 9\\
\hline
\multicolumn{6}{c}{Fe K emission lines}\\
\hline
$6400\pm40$ & $90\pm33$ & $2.8\pm0.9$ & $40\pm15$ & Fe K$\alpha$ & 16\\[+4pt]
$6640\pm30$ & $<250^{a}$ & $1.1\pm0.7$ & $16\pm10$ & Fe~XXV 2s$\rightarrow$2p & 6\\ 
\hline
\multicolumn{6}{c}{Soft X-ray absorption lines}\\
\hline
$922^{+7}_{-3}$ & $1.4^{+0.5}_{-1.2}$ & $-2.3\pm0.1$ & $-1.3\pm0.6$ & Fe~XIX 2p$\rightarrow$3d & 7\\[+4pt]
$944\pm4$ & $1.5^{+1.5}_{-0.7}$ & $-1.6\pm0.9$ & $-1.0\pm0.5$ & Fe~XVIII 2s$\rightarrow$3p & 4\\[+4pt]
$962\pm5$ & $4.0\pm3.0$ & $-2.4\pm1.1$ & $-1.5\pm0.9$ & Fe~XX 2p$\rightarrow$3d & 5\\[+4pt]
$988\pm4$ & $5.0\pm4.0$ & $-2.3\pm1.3$ & $-1.5\pm0.8$ & Fe~XX 2p$\rightarrow$3d & 4\\
\enddata
\tablecomments{Columns: rest-frame energy in eV; line width in eV; intensity in units of $10^{-5}$ ph~s$^{-1}$~cm$^{-2}$; EW in eV; line identification; C-statistics improvement over degrees of freedom $\nu$. $^{a}$ 90\% upper limit.}
\end{deluxetable*}

\subsubsection{Higher resolution view of specific energy intervals}

We performed a test using the highest resolution of the \emph{Chandra} HETG ($\Delta\lambda =5$ m{\AA} and $\Delta\lambda = 10$ m{\AA} for the HEG and MEG, respectively) to check for possible unresolved features and to check line profiles in specific energy bands. In Fig.~3 we show the data binned down to half of the FWHM. We note that the resolution in Fig.~3 is eight times higher than in Fig.~2 (4$\times$FWHM). The limited signal-to-noise does not allow to perform a detailed analysis, although some possible interesting features are discussed in this section. 

Looking at the Fe~L band in the energy interval E$=$0.7--0.8~keV in the upper panel of Fig.~3 it is clear that there is an excess of emission here due to several narrow lines. The main Fe~XVII/XVIII lines are marked in Fig.~3. In the Ne band between E$=$0.900--0.926~keV shown in the second panel in Fig.~3 there could be some possible interesting structures. The absorption at E$\simeq$920~eV is also consistent with the expected position of the resonance absorption of Ne~IX at E$=$920~eV. So, it may be possible that Ne absorption partially contributes to the absorption line interpreted as Fe~XIX in Table~2. 

In the energy band between E$=$1.3--1.5~keV shown in the third panel of Fig.~3 there seems to be a possible emission feature due to the Mg~XII Ly$\alpha$ emission line. The line energy appears to be slightly redshifted at E$=$$1436\pm7$~eV with respect to the expected energy at E$=$1470~eV. The profile is also broadened with a width of $\sigma$$=$$13^{+8}_{-4}$~eV, which would correspond to a FWHM of about 6000~km~s$^{-1}$. We do not find any obvious additional features due to the Si K-shell. 

The Fe K band region shown in the lower panel of Fig.~3 shows a possible narrow core of the Fe K$\alpha$ line at the energy of $E$$\simeq$6.44~keV. The line is unresolved, with a limit of $\sigma$$<$13~eV at the 90\% confidence, which would correspond to a FWHM of less than 1400~km~s$^{-1}$. The line seems faint, and it is significant at a level of $\Delta C = 6$. There seems to be an apparent excess of emission on both red and blue side of the narrow core and the width of $\sigma$$\simeq$90~eV estimated considering only one Gaussian line in Table~2 may be due to the sum of the narrow core and a possible broader component. However, the limited signal-to-noise does not allow further investigation. We note that also the optical H$\alpha$ emission line in this source has a broad, complex double-peaked profile with a narrow core and broad wings (Sambruna et al.~2009).

  \begin{figure}
  \centering
   \includegraphics[width=6.7cm,height=4.8cm,angle=0]{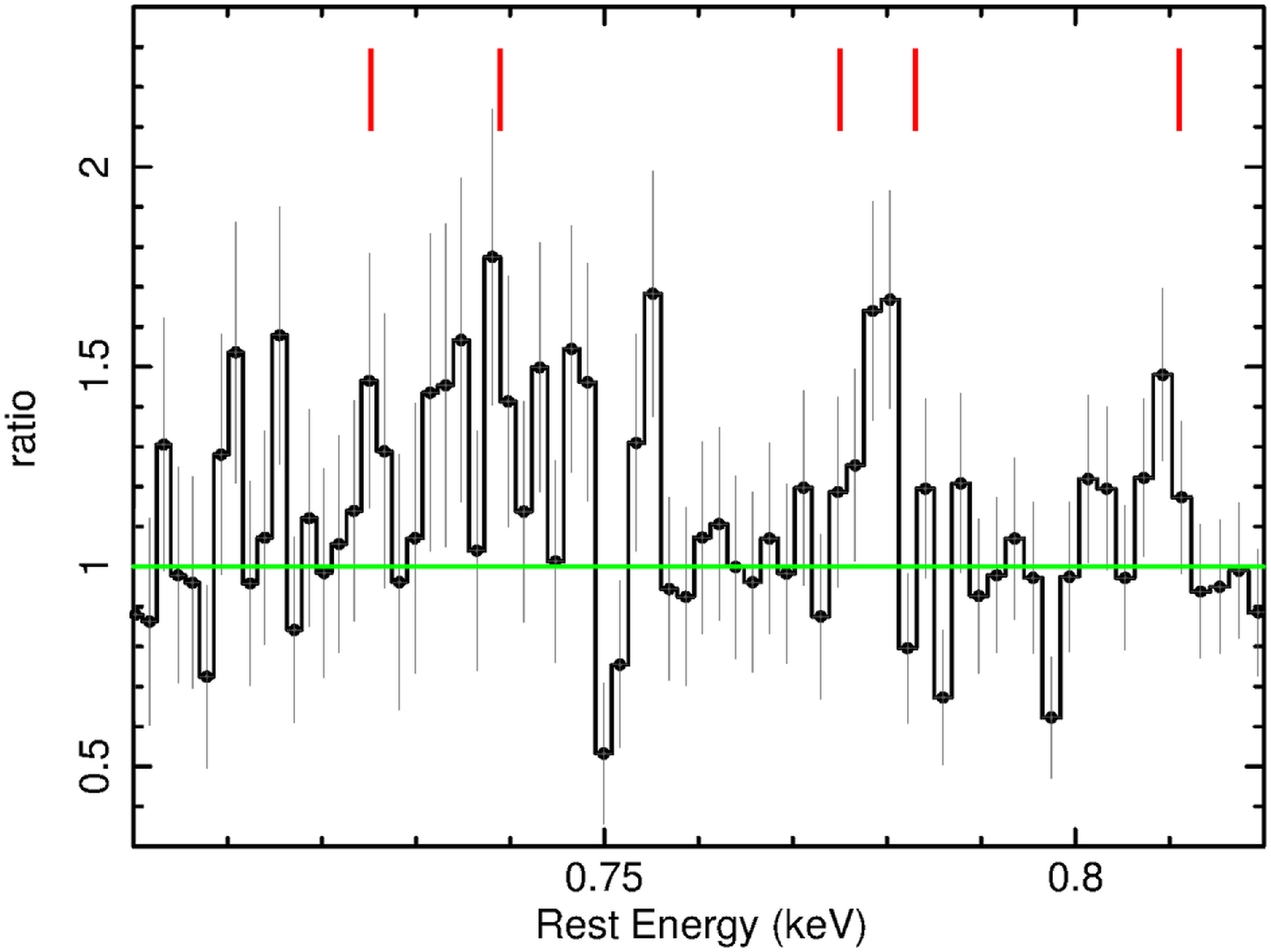}
   \includegraphics[width=6.7cm,height=4.8cm,angle=0]{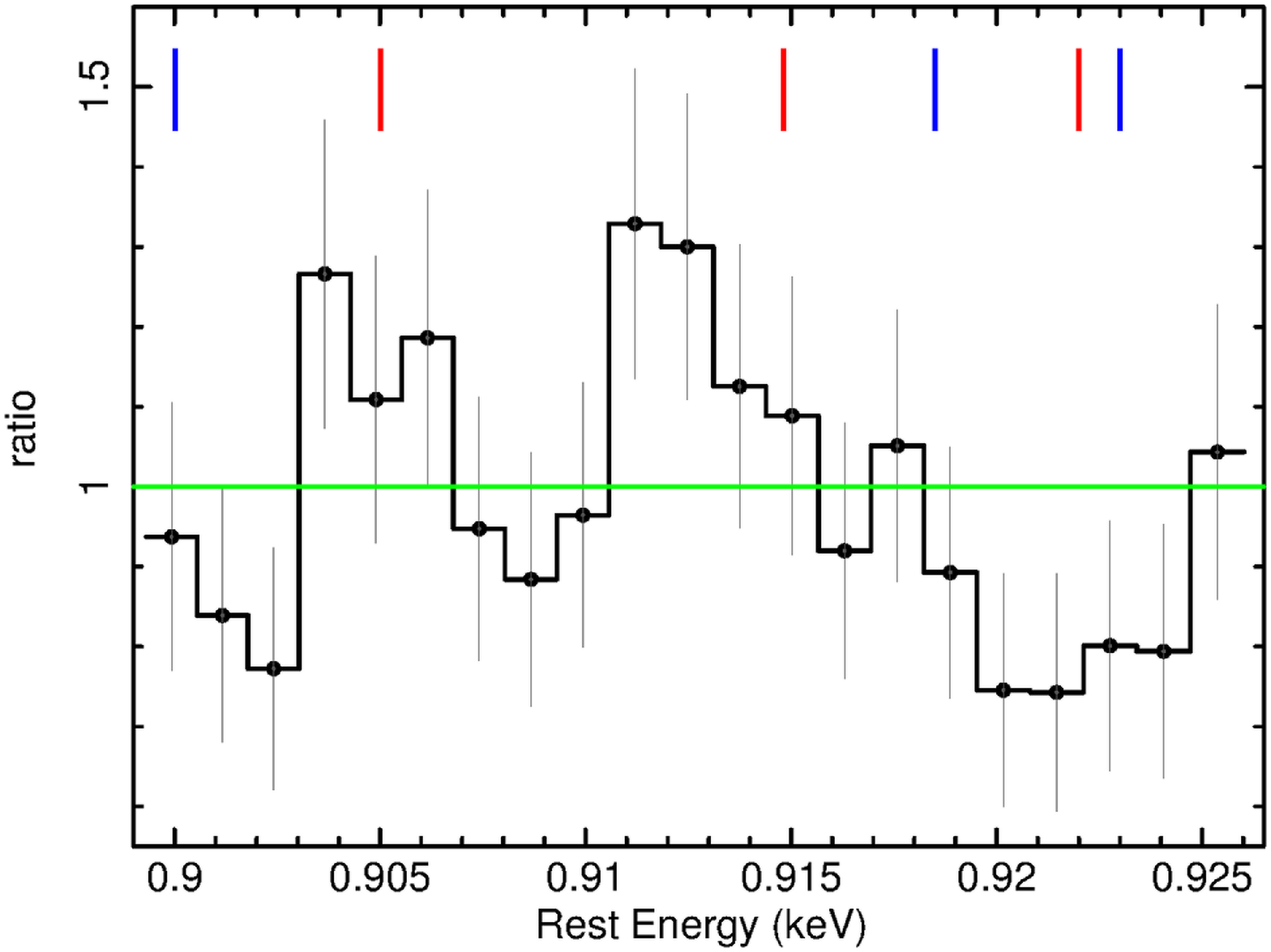}
   \includegraphics[width=6.7cm,height=4.8cm,angle=0]{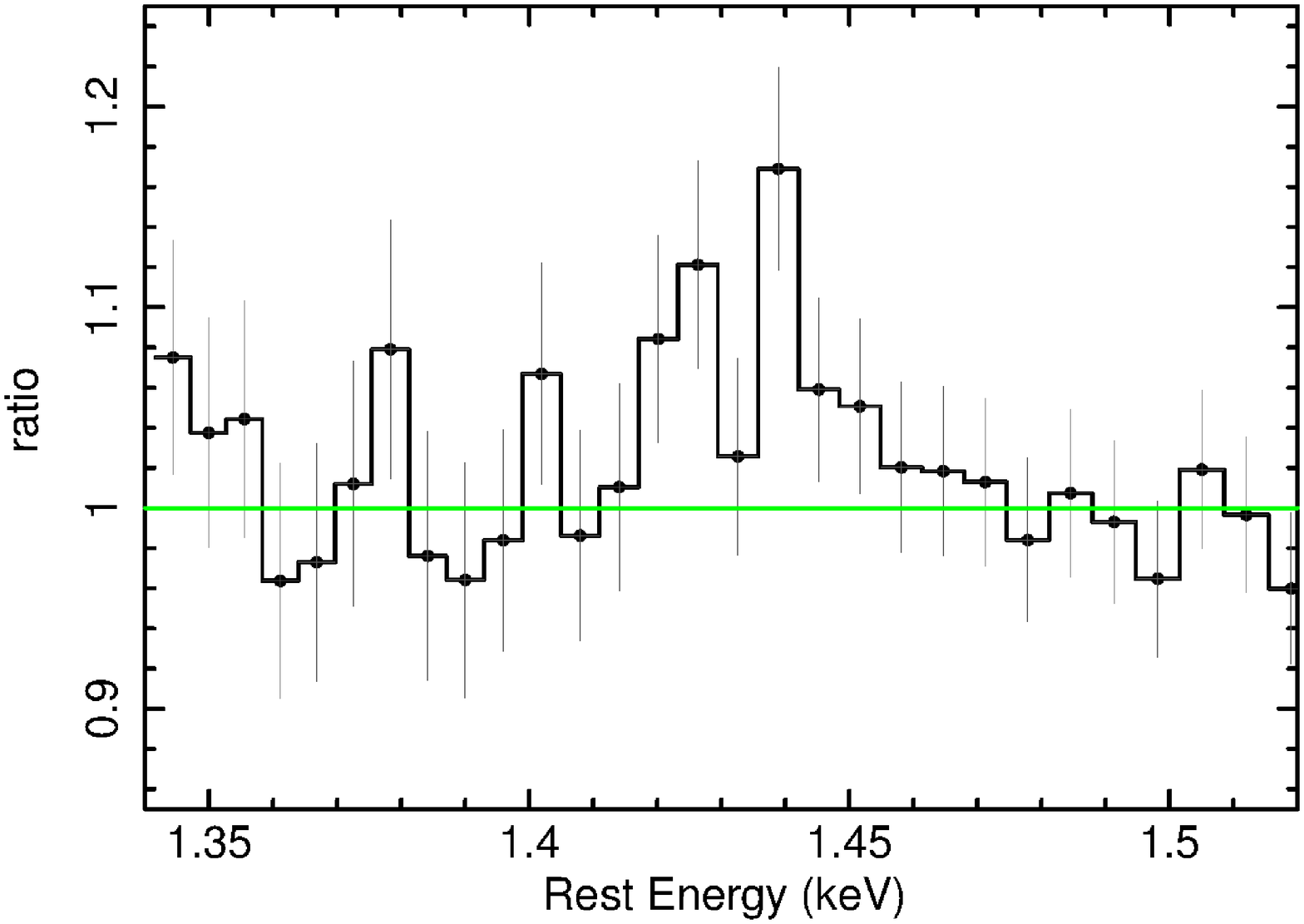}
   \includegraphics[width=6.7cm,height=4.8cm,angle=0]{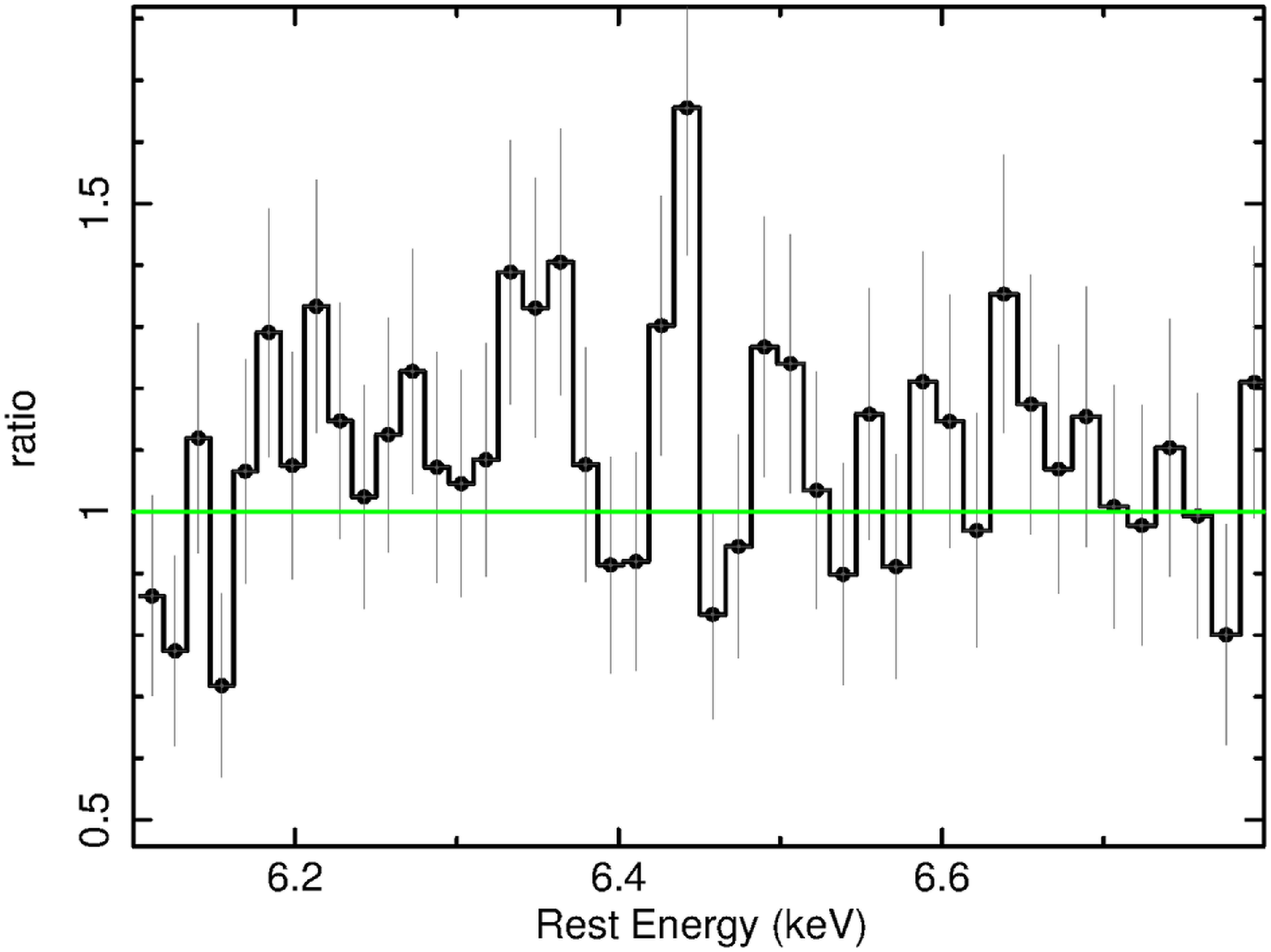}
   \caption{Data to model ratios of the \emph{Chandra} HETG spectrum of 3C~390.3 with respect to a Galactic absorbed power-law continuum model in different energy bands. The MEG data are shown in the first three panels. The HEG data are shown in the fourth panel. The data are binned down to half of the FWHM resolution. The top panel shows the Fe L-shell band, where the positions of possible Fe~XVII lines are marked. The second panel shows region of the Ne~IX triplet (red) and Fe~XIX (blue) with the lines marked. The third panel show a possible broad component of Mg~XII emission and the fourth panel show the high resolution view of the iron K band.}
    \end{figure}

\subsection{Spectral analysis with physical models}

After the initial phenomenological modeling, we performed a more physically motivated fit of the spectral features in the spectrum. 

\subsubsection{Hot emitting gas}

We investigated two possible cases for the emission in the soft X-rays using the photo-ionization model \emph{photemis}\footnote{http://heasarc.gsfc.nasa.gov/docs/software/xstar/xstar.html} (Kallman \& Bautista 2001) and the collisional ionization model \emph{apec} (Smith et al.~2001). We find that the collisional ionization \emph{apec} model provides the best-fit solution with a temperature of $kT$$\simeq$0.5~keV ($\Delta C/\Delta\nu = 10/2$ corresponding to a confidence level of 99.3\%). A comparable fit with \emph{photemis} would require too low abundances ($<$0.01 Solar) of low-Z elements, such as O and Ne, relative to Fe. Instead, the collisional model provides a good fit with abundances consistent with Solar within the uncertainties. Given the same temperature, the collisional ionization case also provides more intense Fe L-shell emission lines with respect to Fe K-shell compared to the photoionization case, because in the latter the X-ray power-law continuum will result in a higher ionization of the gas (e.g., Kallman, Vrtilek \& Kahn 1989; Kallman 1995). 

\subsubsection{Warm absorber}

We compared the parameterization of the soft X-ray absorption features with a photo-ionization model \emph{warmabs} and a collisional ionization model \emph{hotabs} (Kallman \& Bautista 2001). This allows to test the possibility that the emission and absorption originate from the same hot gas (e.g., Sarazin 1989; David 2000) or if the absorption is dominated by the gas photoionized by the AGN. We consider standard Solar abundances. The photo-ionization model provides a much better representation of the absorber compared to the collisional case, i.e. $\Delta C/\Delta\nu$$=$23/2 instead of $\Delta C/\Delta\nu$$=$10/2. The collisional model alone is not able to explain the observed intensity of the Fe L-shell absorption lines. We estimate an ionization parameter log$\xi$$\simeq$2.3~erg~s$^{-1}$~cm and column density log$N_H$$\simeq$20.7~cm$^{-2}$. The confidence level of the photo-ionized absorber is $>$99.99\%. We can only place an upper limit on the outflow velocity of the absorber of $v_{out}$$<$150~km~s$^{-1}$. Equivalent fits were obtained considering a turbulent velocity parameter of the \emph{warmabs} model in the range between 100~km~s$^{-1}$ and 500~km~s$^{-1}$. These values are consistent with the widths of the absorption lines when fitted with inverted Gaussians. The best-fit parameters are reported in Table~3.

\subsubsection{Fe K emission}

We replaced the Fe K$\alpha$ emission line in turn with a cold reflection component \emph{pexmon} (Nandra et al.~2007), an ionized reflection component \emph{xillver} (Garc{\'{\i}}a et al.~2014), and a photo-ionized emission component \emph{photemis} (Kallman \& Bautista 2001). We assumed an inclination angle of 30$^{\circ}$ consistent with the radio jet, a standard Solar abundance for iron and a high energy cut-off of E$=$150~keV as estimated from hard X-ray observations (Sambruna et al.~2009). The best-fit is provided by a lowly ionized, log$\xi$$\simeq$1.3~erg~s$^{-1}$~cm, \emph{xillver} component ($\Delta C/\Delta\nu$$=$14/2). A Gaussian broadening of $\sigma$$\simeq$80~eV is highly required by the data ($\Delta C/\Delta\nu$$=$10/1). The ionized reflection is required with a confidence level $>$99.99\%. The inclusion of a photo-ionized \emph{photemis} component with log$\xi$$\simeq$3.5 erg~s$^{-1}$~cm to model the Fe~XXV emission line is marginally required by the data. Therefore, we will not include this latter component in the final best-fit model reported in Table~3. 

\section{Discussion}

The analysis of the \emph{Chandra} HETG spectrum of 3C~390.3 reveals a complex environment in this radio galaxy. In the following sections we describe the possible physical origin of each component.

\begin{deluxetable}{lc}
\tablecaption{Best-fit model of the \emph{Chandra} HETG spectrum.}
\tablehead
{
\multicolumn{2}{c}{Power-law continuum}
}
\startdata
$\Gamma$  & $1.71\pm0.01$\\
\hline\hline
\multicolumn{2}{c}{gsmooth $\times$ xillver}\\
\hline
$\sigma$ (eV) & $75\pm30$\\
log$\xi$  (erg~s$^{-1}$~cm) & $1.30\pm0.3$\\
\hline\hline
\multicolumn{2}{c}{apec} \\
\hline
$kT$ (keV) & $0.5\pm0.1$\\
$EM$ (cm$^{-3}$) & $5.4\times 10^{64}$ \\ 
\hline\hline
\multicolumn{2}{c}{warmabs} \\
\hline
log$\xi$ (erg~s$^{-1}$~cm) & $2.3\pm0.5$\\
log$N_H$ (cm$^{-2}$) & $20.7\pm0.1$\\
$v_{out}$ (km~s$^{-1}$) & $<$150$^*$\\ 
\hline\hline
\multicolumn{2}{c}{Best-fit}\\
\hline
C/$\nu$ & 9118/8891\\
\hline\hline
\multicolumn{2}{c}{Power-law flux$^a$} \\
\hline
0.5--2~keV  & 3.0\\
2--10~keV   & 5.4\\
\enddata
\tablenotetext{a}{Absorption corrected flux in units of $10^{-11}$ erg~s$^{-1}$~cm$^{-2}$}
\tablenotetext{*}{90\% upper limit.}
\end{deluxetable}

\subsection{Warm absorber}

The soft X-ray band shows the presence of a warm absorber with ionization parameter log$\xi$$\simeq$2.3~erg~s$^{-1}$~cm, column density log$N_H$$\simeq$20.7~cm$^{-2}$ and outflow velocity of $v_{out}$$<$150~km~s$^{-1}$, see Table~3. The parameters of the WA in the 2014 \emph{Chandra} HETG spectrum are consistent with those reported by Torresi et al.~(2012) derived from the 2004 \emph{XMM-Newton} RGS spectrum. Therefore, the WA did not drastically vary on a time-scale of about 10 years, supporting the estimate of a large distance from the central AGN. Consequently, the lower limit of the distance of the absorber from the X-ray source can then be estimated from the light-crossing time argument to be $r_{\mathrm{min}} \simeq $3.5~pc. The upper limit on the location of the absorber can be derived from the definition of the ionization parameter and assuming that the thickness is lower than the distance: $r_{\mathrm{max}} \equiv L_{\mathrm{ion}}/\xi N_\mathrm{H}$ (e.g., Tombesi et al.~2013a). The absorption corrected ionizing luminosity extrapolated between 1--1000~Ryd (1~Ryd$=$13.6~eV) is $L_{\mathrm{ion}}$$\simeq$$1\times 10^{45}$~erg~s$^{-1}$. Substituting the values in Table~3 we obtain an upper limit on the distance of $\sim$3.5~kpc. Therefore, the location of the observed WA is likely of the order of the narrow line region, between $\simeq$3.5~pc and $\simeq$3.5~kpc (e.g., Torresi et al.~2012). 

The mass outflow rate can be estimated from the formula $\dot{M}_{out} = 4\pi \mu m_p r N_H C_F v_{out}$, where $\mu = 1.4$, $m_p$ is the proton mass and $C_F$ the covering fraction (e.g., Crenshaw \& Kraemer 2012; Tombesi et al.~2015). Substituting the values in Table 3, the lower and upper limits in the distance and typical value $C_F\simeq0.5$ (Crenshaw \& Kraemer 2012; Tombesi et al.~2013a, 2014) we obtain the values $\dot{M}_{out}$$\simeq$0.01--15~$M_\odot$~yr$^{-1}$. The mechanical power of the WA can then be estimated as $\dot{E}_K = (1/2)\dot{M}_{out}v_{out}^2$ and we obtain values in the interval between $1\times 10^{38}$~erg~s$^{-1}$ and $1\times 10^{41}$~erg~s$^{-1}$. The estimates of the bolometric luminosity of the AGN and the radio jet power are $5\times 10^{45}$~erg~s$^{-1}$ and $1.3\times 10^{45}$~erg~s$^{-1}$, respectively (Torresi et al.~2012). Therefore, the mechanical power of the WA is negligible, being lower than 0.001\% and 0.01\% of the bolometric luminosity and jet power, respectively. Therefore, the large-scale WA does not seem to provide a significant AGN feedback effect in 3C~390.3, unless it is interpreted as the remnant of the shock interaction between the disk wind and the interstellar medium (e.g., Pounds \& King 2013; King \& Pounds 2014). 

A UFO with velocity $v_{out}$$\simeq$0.15c was detected in 3C~390.3 (Tombesi et al.~2010b, 2014; Gofford et al.~2013, 2015). The lower and upper limits on the distance, mass outflow rate and mechanical power are $d$$\simeq$0.002--0.02~pc, $\dot{M}_{out}$$\simeq$2--12~$M_{\odot}$~yr$^{-1}$ and $\dot{E}_K$$\simeq$$3\times10^{44}$--$4\times10^{45}$~erg~s$^{-1}$, respectively. We note that the mass outflow rate of the UFO is comparable to that estimated for the WA. The mechanical power of the UFO is indeed high enough to provide AGN feedback in this source, with a value comparable to that of the radio jet (Tombesi et al.~2014; Gofford et al.~2015). 

\subsection{Hot interstellar medium}

The soft X-ray emission lines indicate the presence of hot diffuse gas with a temperature of $kT$$\simeq$0.5~keV. This temperature is consistent with the typical value observed for the hot ISM in elliptical galaxies (e.g., Werner et al.~2009; Kim \& Pellegrini 2012). The normalization of the \emph{apec} component is defined as $N = 10^{-14}(EM/4\pi[D_A (1+z)]^2)$, where $D_A \simeq 6.63\times 10^{26}$~cm is the angular diameter distance to the source. The emission measure $EM$ is defined as $EM = \int n_e n_H dV$, where $n_e$ and $n_H$ are the electron and hydrogen number densitites in cm$^{-3}$, respectively. Given the normalization of the \emph{apec} component reported in Table~3, we can estimate an emission measure of $EM \simeq 5.42\times 10^{64}$ cm$^{-3}$. Values of the emission measure of order $10^{64}$ cm$^{-3}$ were reported also for several elliptical galaxies observed with the \emph{XMM-Newton} Reflection Grating Spectrometer (Werner et al.~2009). Assuming that the plasma is fully ionized, $n_e \simeq 1.2 n_H$, and considering a roughly spherical volume of radius $\simeq$1.5kpc corresponding to the extension of the region observed with the HETG of $\simeq$2'', we obtain a number density for the gas of $n_H \sim 0.1$~cm$^{-3}$. The temperature, density and extension of the gas are indeed consistent with the values expected for the hot ISM in elliptical galaxies (Xu et al.~2002; Werner et al.~2009; Kim \& Pellegrini 2012).

Most elliptical and other early-type galaxies possess a hot and diffuse interstellar medium. The ISM temperature ranges from E$\simeq$0.5 to E$\simeq$1~keV (e.g., Matsumoto et al.~1997) so that most of the emission is radiated in the soft X-ray band. Elliptical galaxies, and giant elliptical galaxies in particular, are the most massive and the oldest stellar systems in the universe. 
The central regions of elliptical galaxies should exhibit strong cooling flows (e.g., Nulsen, Stewart, \& Fabian 1984). However, such cooling flows are hardly observed, indicating that the cooling of the gas is counterbalanced by other mechanisms, possibly related to stellar winds, supernovae or AGN wind/jets (e.g., Xu et al.~2002). The cooling flow problem in elliptical galaxies can be regarded as a scaled-down version of the similar process in galaxy clusters. The study of radio galaxies in the X-ray band is therefore an important "laboratory" for cluster physics and AGN feedback at scales of isolated galaxies or galaxy groups. 

The total X-ray luminosity estimated for the hot ISM in 3C~390.3 is $L_{ISM}$$\simeq$$2.5\times 10^{42}$~erg~s$^{-1}$. In order to keep the gas hot and prevent a catastrophic cooling flow, there must be some source of energy. An order of magnitude estimate of the cooling timescale for the hot gas associated with the galaxy can be estimated as $t_{cool} = U/L_{ISM}$, where $U$ is the internal energy of the gas and $L_{ISM}$ is the luminosity. The internal egergy of the gas is defined as $U = (3/2) NkT$, where $N$ is the total number of particles and $T$ is the gas temperature. For a particle density of $n \simeq 0.1$ cm$^{­3}$ and considering a spherical region of radius $\sim$$1.5$~kpc and temperature $T \simeq 10^7$~K, then we obtain $U\simeq10^{56}$ ergs. Considering the luminosity $L_{ISM} \simeq 10^{42}$ ergs~s$^{-1}$ for the hot ISM, then the cooling time is significantly shorter than a Gyr. Even this rough estimate already suggests that an additional source of heating is required to keep the plasma in an approximate equilibrium. 

The AGN at the center of this galaxy has three possible ways to inject energy in the ISM, through the observed warm absorber, the ultrafast outflow and the radio jet. The kinetic energy of these outflows is in the range $\dot{E}_K \simeq 10^{40}$ erg~s$^{-1}$, $\dot{E}_K$$\simeq 10^{44}$ erg~s$^{-1}$, and $\dot{E}_K \simeq 10^{45}$~erg~s$^{-1}$, respectively. The kinetic energy in the warm absorber is at least one order of magnitude below the one required. Instead, even if just $\simeq$1\% of the energy from the ultrafast outflow or the jet are deposited in the ISM, this is enough to counterbalance the cooling. Therefore, it is likely that these two components may provide an important contribution in the balance between heating and cooling in this source.    

Assuming the possible distance of $\sim$1~kpc, then we derive that the density of the warm absorber material, $n\simeq N_H/R \simeq 3\times 10^{20}/3\times 10^{21}$$\sim$0.1 cm$^{-3}$, is comparable to the one derived for the emitting ISM. Therefore, the fact that the emission component is better described by collisional ionization can be probably attributed to the fact that we are looking directly down the ionization cone of this source and that the AGN only photoionizes along the unobscured cone, vertical to the obscuring torus. This is overall consistent with the angle-dependent unified picture of AGN (e.g., Antonucci 1984, 1993; Urry \& Padovani 1995).

\subsection{Fe K band emission}

The most intense emission feature in the Fe K band at E$\simeq$6.4~keV is likely associated with the Fe K$\alpha$ fluorescent line. The best-fit is provided by a lowly ionized  \emph{xillver} component with log$\xi$$\simeq$1.3~erg~s$^{-1}$~cm, see Table~3. The line is resolved and a broadening of $\sigma$$=$$75\pm30$~eV is required. This corresponds to a velocity broadening of FWHM$=$$8300\pm3300$~km~s$^{-1}$. This is consistent with the width of the H$\alpha$ line in the optical, corroborating the previously suggested origin of this feature as reflection from the broad line region or the outer accretion disk (e.g., Sambruna et al.~2009; Zhang 2013). Indeed, we note that also the optical H$\alpha$ emission line in this source has a broad, complex double-peaked profile with a narrow core and broad wings (Sambruna et al.~2009). The signal-to-noise of the data is not enough to significantly investigate the presence of a possible additional broad reflection component associated with the inner accretion disk. 

We find marginal evidence for a Fe~XXV emission line indicating the presence of an additional higher ionization component with ionization parameter  log$\xi$$\simeq$3.5 erg~s$^{-1}$~cm. Increasing evidence for such ionized Fe K emission lines has been reported in the spectra of both radio-quiet and radio-loud AGN (e.g., Bianchi et al.~2005; Tombesi et al.~2010a, b; Patrick et al.~2012; Lohfink et al.~2012; Tombesi et al.~2013). The origin of such features is still unclear, possibly being associated with disk winds, extended photoionied gas or disk reflection. Longer exposure observations with \emph{Chandra} and \emph{XMM-Newton} will help shed light on these features. A fundamental improvement in the study such narrow Fe K features will be provided by future X-ray calorimeter missions (e.g., Nandra et al.~2013; Takahashi et al.~2014).    

\section{Conclusions}

We report for the first time the analysis of the 150ks \emph{Chandra} HETG spectrum of the radio galaxy 3C~390.3. The spectrum shows complex emission and absorption features in both the soft X-rays and the Fe K band. In particular, there is evidence for (i) a hot emission component likely associated with the hot ISM in this elliptical galaxy, (ii) a warm absorber located in the range between 0.25--3.5~kpc, and (iii) lowly ionized reflection in the Fe K band likely associated with the optical broad line region or the outer accretion disk, with a hint of a higher ionization component. This is the first case in which these three components have been simultaneously observed in a normal elliptical galaxy hosting a luminous AGN with a powerful relativistic jet and accretion disk winds.

\acknowledgments

F.T. thanks K. Fukumura, D. Kazanas and F. Paerels for the useful discussions. F.T. acknowledges support for this work by the National Aeronautics and Space Administration (NASA) through Chandra Award Number GO4-15103A issued by the Chandra X-ray Observatory Center, which is operated by the Smithsonian Astrophysical Observatory for and on behalf of NASA under contract NAS8-03060. E.B. received funding from the EU Horizon 2020 research and innovation programme under the Marie Sklodowska­Curie grant agreement no. 655324.


\begin{thebibliography}{}
\bibitem[Antonucci(1984)]{1984ApJ...278..499A} Antonucci, R.~R.~J.\ 1984, ApJ, 278, 499 
\bibitem[Antonucci(1993)]{1993ARA&A..31..473A} Antonucci, R.\ 1993, ARA\&A, 31, 473 
\bibitem[Ballantyne(2005)]{2005MNRAS.362.1183B} Ballantyne, D.~R.\ 2005, MNRAS, 362, 1183 
\bibitem[Ballo et al.(2011)]{2011MNRAS.418.2367B} Ballo, L., Braito, V., Reeves, J.~N., Sambruna, R.~M., \& Tombesi, F.\ 2011, MNRAS, 418, 2367 
\bibitem[Bianchi et al.(2005)]{2005MNRAS.357..599B} Bianchi, S., Matt, G., Nicastro, F., Porquet, D., \& Dubau, J.\ 2005, MNRAS, 357, 599 
\bibitem[Blandford \& Payne(1982)]{1982MNRAS.199..883B} Blandford, R.~D., \& Payne, D.~G.\ 1982, MNRAS, 199, 883 
\bibitem[Bostrom et al.(2014)]{2014ApJ...791..119B} Bostrom, A., Reynolds, C.~S., \& Tombesi, F.\ 2014, ApJ, 791, 119 
\bibitem[Braito et al.(2011)]{2011MNRAS.414.2739B} Braito, V., Reeves, J.~N., Sambruna, R.~M., \& Gofford, J.\ 2011, MNRAS, 414, 2739 
\bibitem[Chatterjee et al.(2009)]{2009ApJ...704.1689C} Chatterjee, R., Marscher, A.~P., Jorstad, S.~G., et al.\ 2009, ApJ, 704, 1689 
\bibitem[Chatterjee et al.(2011)]{2011ApJ...734...43C} Chatterjee, R., Marscher, A.~P., Jorstad, S.~G., et al.\ 2011, ApJ, 734, 43 
\bibitem[Chiaberge \& Marconi(2011)]{2011MNRAS.416..917C} Chiaberge, M., \& Marconi, A.\ 2011, MNRAS, 416, 917 
\bibitem[Chiaberge et al.(2015)]{2015ApJ...806..147C} Chiaberge, M., Gilli, R., Lotz, J.~M., \& Norman, C.\ 2015, ApJ, 806, 147 
\bibitem[Crenshaw & Kraemer(2012)]{2012ApJ...753...75C} Crenshaw, D.~M., \& Kraemer, S.~B.\ 2012, ApJ, 753, 75 
\bibitem[David(2000)]{2000ApJ...529..682D} David, L.~P.\ 2000, ApJ, 529, 682 
\bibitem[Fabian(1999)]{1999MNRAS.308L..39F} Fabian, A.~C.\ 1999, MNRAS, 308, L39 
\bibitem[Fabian(2012)]{2012ARA&A..50..455F} Fabian, A.~C.\ 2012, ARA\&A, 50, 455 
\bibitem[Fukumura et al.(2010)]{2010ApJ...715..636F} Fukumura, K., Kazanas, D., Contopoulos, I., \& Behar, E.\ 2010, ApJ, 715, 636
\bibitem[Fukumura et al.(2014)]{2014ApJ...780..120F} Fukumura, K., Tombesi, F., Kazanas, D., et al.\ 2014, ApJ, 780, 120 
\bibitem[Fukumura et al.(2015)]{2015ApJ...805...17F} Fukumura, K., Tombesi, F., Kazanas, D., et al.\ 2015, ApJ, 805, 17  
\bibitem[Garc{\'{\i}}a et al.(2014)]{2014ApJ...782...76G} Garc{\'{\i}}a, J., Dauser, T., Lohfink, A., et al.\ 2014, ApJ, 782, 76 
\bibitem[Gofford et al.(2013)]{2013MNRAS.430...60G} Gofford, J., Reeves, J.~N., Tombesi, F., et al.\ 2013, MNRAS, 430, 60 
\bibitem[Gofford et al.(2015)]{2015MNRAS.451.4169G} Gofford, J., Reeves, J.~N., McLaughlin, D.~E., et al.\ 2015, MNRAS, 451, 4169 
\bibitem[Grandi \& Palumbo(2007)]{2007ApJ...659..235G} Grandi, P., \& Palumbo, G.~G.~C.\ 2007, ApJ, 659, 235 
\bibitem[Halpern(1984)]{1984ApJ...281...90H} Halpern, J.~P.\ 1984, ApJ, 281, 90 
\bibitem[Ineson et al.(2015)]{2015MNRAS.453.2682I} Ineson, J., Croston, J.~H., Hardcastle, M.~J., et al.\ 2015, MNRAS, 453, 2682 
\bibitem[Kaastra et al.(2014)]{2014Sci...345...64K} Kaastra, J.~S., Kriss, G.~A., Cappi, M., et al.\ 2014, Science, 345, 64 
\bibitem[Kalberla et al.(2005)]{2005A&A...440..775K} Kalberla, P.~M.~W., et al.\ 2005, A\&A, 440, 775
\bibitem[Kallman et al.(1989)]{1989ApJ...345..498K} Kallman, T.~R., Vrtilek, S.~D., \& Kahn, S.~M.\ 1989, ApJ, 345, 498 
\bibitem[Kallman(1995)]{1995ApJ...455..603K} Kallman, T.~R.\ 1995, ApJ, 455, 603 
\bibitem[Kallman \& Bautista(2001)]{2001ApJS..133..221K} Kallman, T., \& Bautista, M.\ 2001, ApJS, 133, 221 
\bibitem[Kataoka et al.(2007)]{2007PASJ...59..279K} Kataoka, J., Reeves, J.~N., Iwasawa, K., et al.\ 2007, PASJ, 59, 279 
\bibitem[Kim & Pellegrini(2012)]{2012ASSL..378.....K} Kim, D.-W., \& Pellegrini, S.\ 2012, Astrophysics and Space Science Library, 378
\bibitem[King \& Pounds(2014)]{2014MNRAS.437L..81K} King, A.~R., \& Pounds, K.~A.\ 2014, MNRAS, 437, L81 
\bibitem[King & Pounds(2015)]{2015ARA&A..53..115K} King, A., \& Pounds, K.\ 2015, ARA\&A, 53, 115 
\bibitem[Kinkhabwala et al.(2002)]{2002ApJ...575..732K} Kinkhabwala, A., Sako, M., Behar, E., et al.\ 2002, ApJ, 575, 732 
\bibitem[Liedahl \& Paerels(1996)]{1996ApJ...468L..33L} Liedahl, D.~A., \& Paerels, F.\ 1996, ApJ, 468, L33 
\bibitem[Lohfink et al.(2013)]{2013ApJ...772...83L} Lohfink, A.~M., Reynolds, C.~S., Jorstad, S.~G., et al.\ 2013, ApJ, 772, 83 
\bibitem[Marscher et al.(2002)]{2002Natur.417..625M} Marscher, A.~P., Jorstad, S.~G., G{\'o}mez, J.-L., et al.\ 2002, Nature, 417, 625 
\bibitem[Matsumoto et al.(1997)]{1997ApJ...482..133M} Matsumoto, H., Koyama, K., Awaki, H., et al.\ 1997, ApJ, 482, 133 
\bibitem[McKinney(2006)]{2006MNRAS.368.1561M} McKinney, J.~C.\ 2006, MNRAS, 368, 1561 
\bibitem[Nardini et al.(2015)]{2015Sci...347..860N} Nardini, E., Reeves, J.~N., Gofford, J., et al.\ 2015, Science, 347, 860 
\bibitem[Nandra \& Pounds(1992)]{1992Natur.359..215N} Nandra, K., \& Pounds, K.~A.\ 1992, Nature, 359, 215 
\bibitem[Nandra et al.(2007)]{2007MNRAS.382..194N} Nandra, K., O'Neill, P.~M., George, I.~M., \& Reeves, J.~N.\ 2007, MNRAS, 382, 194 
\bibitem[Nandra et al.(2013)]{2013arXiv1306.2307N} Nandra, K., Barret, D., Barcons, X., et al.\ 2013, arXiv:1306.2307 
\bibitem[Nulsen et al.(1984)]{1984MNRAS.208..185N} Nulsen, P.~E.~J., Stewart, G.~C., \& Fabian, A.~C.\ 1984, MNRAS, 208, 185 
\bibitem[Patrick et al.(2012)]{2012MNRAS.426.2522P} Patrick, A.~R., Reeves, J.~N., Porquet, D., et al.\ 2012, MNRAS, 426, 2522 
\bibitem[Pounds \& King(2013)]{2013MNRAS.433.1369P} Pounds, K.~A., \& King, A.~R.\ 2013, MNRAS, 433, 1369 
\bibitem[Reeves et al.(2009)]{2009ApJ...702L.187R} Reeves, J.~N., Sambruna, R.~M., Braito, V., \& Eracleous, M.\ 2009, ApJ, 702, L187 
\bibitem[Reeves et al.(2010)]{2010ApJ...725..803R} Reeves, J.~N., Gofford, J., Braito, V., \& Sambruna, R.\ 2010, ApJ, 725, 803 
\bibitem[Reynolds(1997)]{1997MNRAS.286..513R} Reynolds, C.~S.\ 1997, MNRAS, 286, 513 
\bibitem[Reynolds et al.(2015)]{2015ApJ...808..154R} Reynolds, C.~S., Lohfink, A.~M., Ogle, P.~M., et al.\ 2015, ApJ, 808, 154 
\bibitem[Sako et al.(2002)]{2002xsac.conf..191S} Sako, M., Kinkhabwala, A., Kahn, S.~M., et al.\ 2002, X-ray Spectroscopy of AGN with Chandra and XMM-Newton, 191 
\bibitem[Sambruna et al.(2009)]{2009ApJ...700.1473S} Sambruna, R.~M., Reeves, J.~N., Braito, V., et al.\ 2009, ApJ, 700, 1473 
\bibitem[Sarazin(1989)]{1989ApJ...345...12S} Sarazin, C.~L.\ 1989, ApJ, 345, 12 
\bibitem[Silk \& Rees(1998)]{1998A&A...331L...1S} Silk, J., \& Rees, M.~J.\ 1998, A\&A, 331, L1 
\bibitem[Smith et al.(2001)]{2001ApJ...556L..91S} Smith, R.~K., Brickhouse, N.~S., Liedahl, D.~A., \& Raymond, J.~C.\ 2001, ApJ, 556, L91 
\bibitem[Takahashi et al.(2014)]{2014SPIE.9144E..25T} Takahashi, T., Mitsuda, K., Kelley, R., et al.\ 2014, ProcSPIE, 9144, 914425 
\bibitem[Tarter et al.(1969)]{1969ApJ...156..943T} Tarter, C.~B., Tucker, W.~H., Salpeter, E.~E.\ 1969, ApJ, 156, 943
\bibitem[Tazaki et al.(2013)]{2013ApJ...772...38T} Tazaki, F., Ueda, Y., Terashima, Y., Mushotzky, R.~F., \& Tombesi, F.\ 2013, ApJ, 772, 38 
\bibitem[Tombesi et al.(2010)]{2010A&A...521A..57T} Tombesi, F., Cappi, M., Reeves, J.~N., et al.\ 2010a, A\&A, 521, A57
\bibitem[Tombesi et al.(2010)]{2010ApJ...719..700T} Tombesi, F., Sambruna, R.~M., Reeves, J.~N., et al.\ 2010b, ApJ, 719, 700  
\bibitem[Tombesi et al.(2011)]{2011MNRAS.418L..89T} Tombesi, F., Sambruna, R.~M., Reeves, J.~N., Reynolds, C.~S., \& Braito, V.\ 2011, MNRAS, 418, L89
\bibitem[Tombesi et al.(2012)]{2012MNRAS.424..754T} Tombesi, F., Sambruna, R.~M., Marscher, A.~P., et al.\ 2012, MNRAS, 424, 754 
\bibitem[Tombesi et al.(2013)]{2013MNRAS.430.1102T} Tombesi, F., Cappi, M., Reeves, J.~N., et al.\ 2013a, MNRAS, 430, 1102 
\bibitem[Tombesi et al.(2013)]{2013MNRAS.434.2707T} Tombesi, F., Reeves, J.~N., Reynolds, C.~S., Garc{\'{\i}}a, J., \& Lohfink, A.\ 2013b, MNRAS, 434, 2707 
\bibitem[Tombesi et al.(2014)]{2014MNRAS.443.2154T} Tombesi, F., Tazaki, F., Mushotzky, R.~F., et al.\ 2014, MNRAS, 443, 2154 
\bibitem[Tombesi et al.(2015)]{2015Natur.519..436T} Tombesi, F., Mel{\'e}ndez, M., Veilleux, S., et al.\ 2015, Nature, 519, 436 
\bibitem[Torresi et al.(2010)]{2010MNRAS.401L..10T} Torresi, E., Grandi, P., Longinotti, A.~L., et al.\ 2010, MNRAS, 401, L10 
\bibitem[Torresi et al.(2012)]{2012MNRAS.419..321T} Torresi, E., Grandi, P., Costantini, E., \& Palumbo, G.~G.~C.\ 2012, MNRAS, 419, 321 
\bibitem[Urry \& Padovani(1995)]{1995PASP..107..803U} Urry, C.~M., \& Padovani, P.\ 1995, PASP, 107, 803 
\bibitem[Veilleux et al.(2013)]{2013ApJ...776...27V} Veilleux, S., Mel{\'e}ndez, M., Sturm, E., et al.\ 2013, ApJ, 776, 27
\bibitem[Werner et al.(2009)]{2009MNRAS.398...23W} Werner, N., Zhuravleva, I., Churazov, E., et al.\ 2009, MNRAS, 398, 23 
\bibitem[Xu et al.(2002)]{2002ApJ...579..600X} Xu, H., Kahn, S.~M., Peterson, J.~R., et al.\ 2002, ApJ, 579, 600 
\bibitem[Zhang(2013)]{2013MNRAS.431L.112Z} Zhang, X.-G.\ 2013, MNRAS, 431, L112 
\end{thebibliography}
\end{document}